\documentclass[sigconf,nonacm]{acmart}
\usepackage{url}
\usepackage{booktabs}
\usepackage{graphicx}
\usepackage{tikz}
\usetikzlibrary{arrows.meta, positioning}
\usepackage{fontawesome5}
\usepackage{todonotes}
\usepackage{stfloats}
\usepackage{tcolorbox}
\tcbuselibrary{breakable}
\usepackage{xspace}
\usepackage{xspace}

\AtBeginDocument{%
  }

\newcommand{\tool}{\textsc{Pulla}\xspace}

\begin{document}

\title{Pulla: A Parsons Problem Tool for Fine-Grained Behavioral Tracing and Instructor-Facing Problem-Solving Analysis}

\author{Daniel Prol}
\orcid{0009-0006-6522-4322}
\affiliation{
    \institution{University of Houston}
    \city{Houston}
    \state{Texas}
    \country{USA}
}
\email{dprol@uh.edu}

\author{Juho Leinonen}
\orcid{0000-0001-6829-9449}
\affiliation{
  \institution{Aalto University}
  \city{Espoo}
  \country{Finland}
}
\email{juho.2.leinonen@aalto.fi}

\author{Arto Hellas}
\orcid{0000-0001-6502-209X}
\affiliation{
  \institution{Aalto University}
  \city{Espoo}
  \country{Finland}
}
\email{arto.hellas@aalto.fi}

\author{Saleh Alkhamees}
\orcid{0000-0002-7447-5899}
\affiliation{%
  \institution{University of Houston}
  \city{Houston}
  \state{Texas}
  \country{USA}
  }
\email{saalkham@cougarnet.uh.edu}

\author{Amin Alipour}
\orcid{0009-0008-1935-0605}
\affiliation{
    \institution{University of Houston}
    \city{Houston}
    \state{Texas}
    \country{USA}
}
\email{maalipou@central.uh.edu}

\renewcommand{\shortauthors}{Prol et al.}

\begin{abstract}

Existing Parsons problem tools primarily focus on correctness, indicating whether a student solved a problem, but providing limited visibility into the underlying problem-solving process. We address this gap by introducing \tool, a Parsons problem tool that instruments programming assignments to capture fine-grained interaction data. These behavioral traces allow the system to surface recurring difficulty patterns, giving instructors actionable insights to inform targeted intervention decisions.

This paper describes our experience in developing and deploying \tool. We deployed the tool in two university courses: an upper-division software
design course at the University of Houston (United States) and an introductory programming course at Aalto University (Finland). By analyzing the data collected, we identified common difficulty patterns, including misidentifying exception types, confusing return with the throw/raise mechanism, and incorrect control-flow ordering.

\end{abstract}





\usetikzlibrary{arrows.meta, positioning}
\definecolor{laneYellow}{HTML}{FEFBE8}
\definecolor{laneBlue}{HTML}{F0F4FF}
\maketitle 
\raggedbottom

\section{Introduction}

Parsons problems are exercises in which students reconstruct a program by arranging shuffled code blocks into the correct order~\cite{parsons2006}. They can optionally include distractors that must be excluded~\cite{parsons2006, ericson2022parsons}. 
They naturally generate behavioral traces as students solve them: each block placement, each erase-and-retry cycle, and each intermediate solution state encodes conceptual decisions that a student makes during problem solving.
While Parsons problems have been extended in many directions (interaction traces~\cite{helminen2012}, adaptive difficulty~\cite{ericson2018adaptive}, faded blocks~\cite{weinman2021faded}, and mobile accessibility~\cite{kluthe2026}) prior work has not provided instructor-facing tools derived from those traces, leaving the problem-solving process invisible to instructors. Without visibility into which blocks caused repeated confusion and which distractors were consistently selected, instructors cannot identify which concepts deserve revisiting or which students may benefit from additional support. To address this gap, we extended an existing Parsons problem tool with behavioral logging and an instructor-facing dashboard, together forming \tool, that visualizes and presents per-student attempt histories and class-level difficulty patterns. The system instruments Parsons assignments to collect fine-grained interaction traces including: complete sequences of intermediate solution states between first block placement and final submission. These behavioral traces are designed to enable instructors to identify recurring difficulty patterns and intervene before those patterns compound into deeper misconceptions. \tool was deployed in two university courses across different institutions and countries, the United States and Finland, over spring and summer semesters, serving real students in graded assignments. 
This paper makes two primary contributions:

\begin{enumerate}
    \item We present a system that records each block arrangement and its per-block color feedback, making each student's complete attempt history and activity logs available to instructors. An AI-assisted pipeline surfaces per-student and class-level difficulty patterns, with instructors remaining the final decision-makers for any intervention.

    \item We report our experience deploying the system across two universities in two countries in real-world classroom settings, providing empirical insights into student behavior and identifying three recurring difficulty patterns.
\end{enumerate}

\section{Related Work}
\label{sec:related}

\paragraph{Parsons Problems as a Formative Assessment Tool.}
Parsons problems have been studied extensively across contexts and skill levels~\cite{du2020review}. Several platforms support their delivery at scale: the js-parsons library\footnote{\url{https://js-parsons.github.io/}} powers web-based courses and was used by Helminen et al.~\cite{helminen2012} to capture interaction traces in Python problems; Runestone~\cite{ericson2020runestone} embeds them in interactive textbooks with adaptive scaffolding; Epplets~\cite{kumar2018epplets} offers credit based on actions taken to reach a solution; and MobileParsons~\cite{karavirta2012mobileparsons} extends the format to iOS and Android. A recurring design choice is the use of distractors: Harms et~al.~\cite{harms2016distractors} and Smith et~al.~\cite{smith2023distractors} found they increase time-on-task by 14--50\% and reduce success rates, demonstrating that distractor presence significantly shapes the solving process. More recently, CodeTailor~\cite{hou2024codetailor} uses an LLM to generate personalized puzzles from a student's own incorrect submission. The format has also been extended to support adaptive difficulty~\cite{ericson2018adaptive}, faded blocks~\cite{weinman2021faded}, and accessible mobile delivery~\cite{kluthe2026}. Despite these advances, existing platforms report correctness and attempt counts but provide no built-in mechanism for instructors to diagnose how students arrive at their solutions, in particular, which distractors they select, how often they backtrack, and what difficulties those choices reveal. Our tool addresses this gap with per-block interaction traces, aggregated activity logs, and on-demand AI-assisted analysis of difficulty patterns.

\paragraph{Process Data and Instructor-Facing Feedback.}
Helminen et al.~\cite{helminen2012} demonstrated the diagnostic value of interaction traces in Parsons problems, revealing patterns such as backtracking and block cycling that correctness scores cannot expose. Price et al. extended this to block-based programming through iSnap~\cite{price2017isnap} and later proposed ProgSnap2~\cite{price2020progsnap2} as a cross-institutional standard for recording programming events from first interaction to final
submission. Benedetti et al.~\cite{Benedetti2026TheUO} illustrate the cost of this gap from a different angle: analyzing written student submissions on algorithmic tasks, they find that when submissions lack explicit solution steps, the reasoning process that produced them remains
opaque to analysis. Despite these advances, process data has remained a research instrument rather than an instructor-facing resource; our tool addresses this gap by exposing each student's complete block-level interaction history within the instructor dashboard.

\paragraph{AI-Assisted Analysis of Student Work.}
Phung et al.~\cite{phung2025} found that expert rubric ratings of AI-generated programming hints tend to overestimate student-perceived usefulness, with most mismatches arising when experts rated a hint as high-quality but students found it unhelpful. A subsequent deployment by the same group~\cite{phung2026} integrated AI-generated hints with an escalation mechanism allowing students to request human instructor feedback when AI support fell short; their results showed that instructor feedback was incorrect or insufficient roughly half the time, even among instructors with domain expertise. S\"{o}lch and Krusche~\cite{soelch2026icse} integrated LLM-based assessment into a production learning management system, generating rubric-aligned feedback on student UML submissions for instructor review before delivery. Recent work has shown that students rate AI-generated lecture slides comparably to human-authored ones and cannot reliably identify which slides are AI-generated ~\cite{leinonen2026aigeneratedslidesgoodstudents}. These approaches take either a student artifact (code, UML) or course content as input and process it without access to the student's problem-solving process. Our approach instead takes a behavioral trace as input and addresses its analysis to the instructor, who remains the decision-maker for any intervention.

\section{System Overview}
\label{sec:system}

\begin{figure*}[t]
  \centering
  \begin{tikzpicture}[
    arr/.style={-{Stealth[scale=0.9]}, thick, color=black},
    panel/.style={draw=black!60, thin, fill=white, rounded corners=2pt, inner xsep=4pt, inner ysep=0pt},
  ]
\node[font=\normalsize, rotate=90] at (-0.35, 3.5)
  {\faUserGraduate~{\small\bfseries Student}};
\node[font=\normalsize, rotate=90] at (-0.35, -4.0)
  {\faChalkboardTeacher~{\small\bfseries Instructor}};
  \node[panel] (ps1) at (2.3,3.9) {\begin{minipage}[c]{3.8cm}\noindent\hspace{-4pt}{\setlength{\fboxsep}{0pt}\colorbox{black}{\parbox{\dimexpr\linewidth+8pt\relax}{\vspace{3pt}\hspace{4pt}\color{white}\scriptsize\bfseries A -- PROBLEM STATEMENT\vspace{3pt}}}}\\[3pt]\includegraphics[width=3.6cm,keepaspectratio]{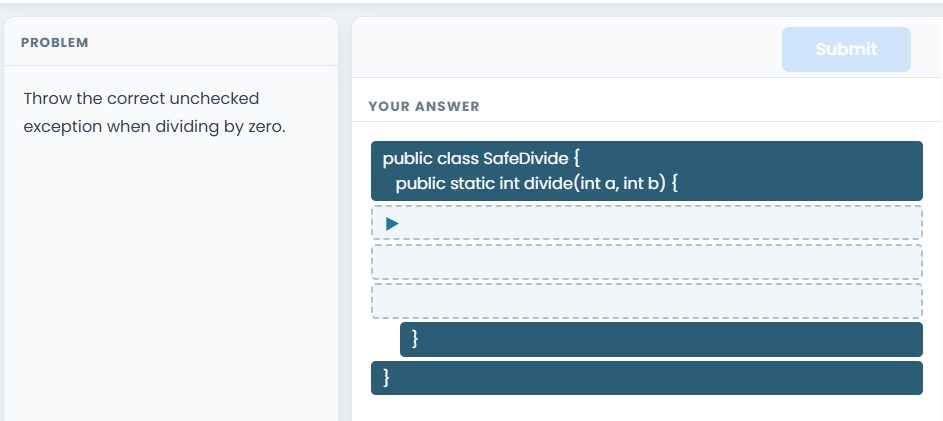}\vspace{3pt}\end{minipage}};%
  \node[panel,right=0.5cm of ps1] (ps2) {\begin{minipage}[c]{3.8cm}\noindent\hspace{-4pt}{\setlength{\fboxsep}{0pt}\colorbox{black}{\parbox{\dimexpr\linewidth+8pt\relax}{\vspace{3pt}\hspace{4pt}\color{white}\scriptsize\bfseries B -- BLOCK BANK\vspace{3pt}}}}\\[3pt]\includegraphics[width=3.6cm,keepaspectratio]{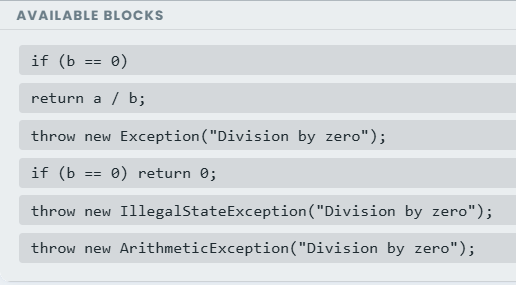}\vspace{3pt}\end{minipage}};%
  \node[panel,right=0.5cm of ps2] (ps3) {\begin{minipage}[c]{3.8cm}\noindent\hspace{-4pt}{\setlength{\fboxsep}{0pt}\colorbox{black}{\parbox{\dimexpr\linewidth+8pt\relax}{\vspace{3pt}\hspace{4pt}\color{white}\scriptsize\bfseries C -- SOLUTION AREA\vspace{3pt}}}}\\[3pt]\includegraphics[width=3.6cm,keepaspectratio]{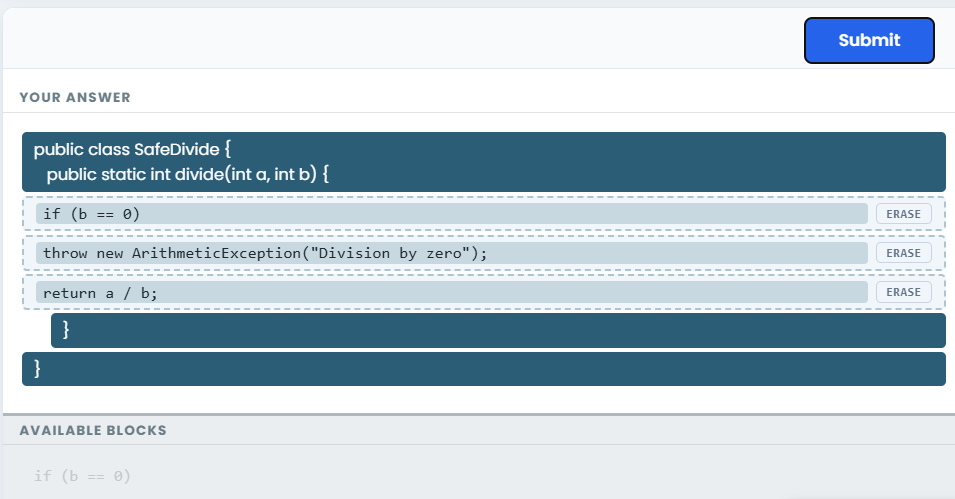}\vspace{3pt}\end{minipage}};%
  \draw[arr] (ps1.east) -- (ps2.west);%
  \draw[arr] (ps2.east) -- (ps3.west);%
  \node[panel] (ps4) at (2.3,0.5) {\begin{minipage}[c]{3.8cm}\noindent\hspace{-4pt}{\setlength{\fboxsep}{0pt}\colorbox{black}{\parbox{\dimexpr\linewidth+8pt\relax}{\vspace{3pt}\hspace{4pt}\color{white}\scriptsize\bfseries D -- FEEDBACK\vspace{3pt}}}}\\[3pt]\includegraphics[width=3.6cm,keepaspectratio]{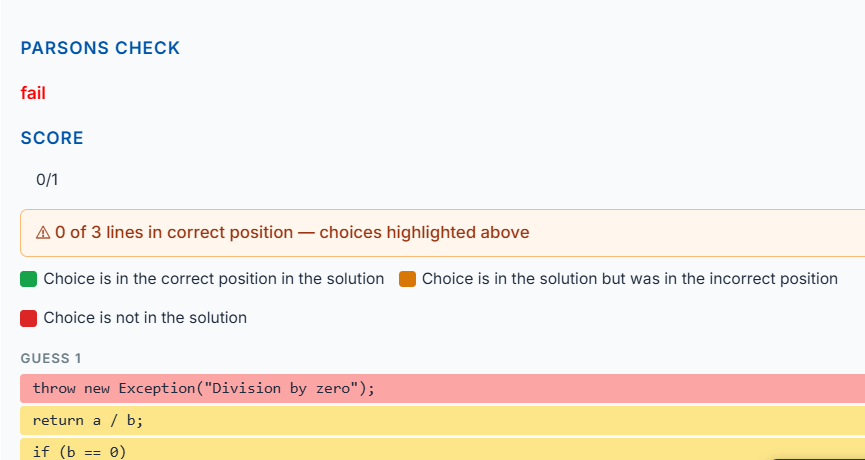}\vspace{3pt}\end{minipage}};%
  \node[panel,right=0.5cm of ps4] (ps5) {\begin{minipage}[c]{3.8cm}\noindent\hspace{-4pt}{\setlength{\fboxsep}{0pt}\colorbox{black}{\parbox{\dimexpr\linewidth+8pt\relax}{\vspace{3pt}\hspace{4pt}\color{white}\scriptsize\bfseries E -- REVISED ATTEMPT\vspace{3pt}}}}\\[3pt]\includegraphics[width=3.6cm,keepaspectratio]{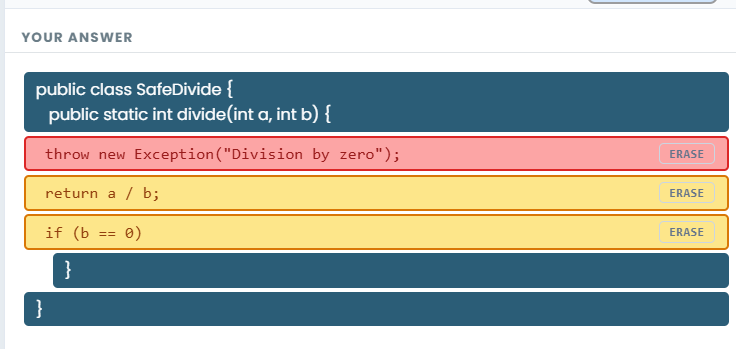}\vspace{3pt}\end{minipage}};%
  \node[panel,right=0.5cm of ps5] (ps6) {\begin{minipage}[c]{3.8cm}\noindent\hspace{-4pt}{\setlength{\fboxsep}{0pt}\colorbox{black}{\parbox{\dimexpr\linewidth+8pt\relax}{\vspace{3pt}\hspace{4pt}\color{white}\scriptsize\bfseries F -- FINAL SUBMISSION\vspace{3pt}}}}\\[3pt]\includegraphics[width=3.6cm,height=2.2cm,keepaspectratio]{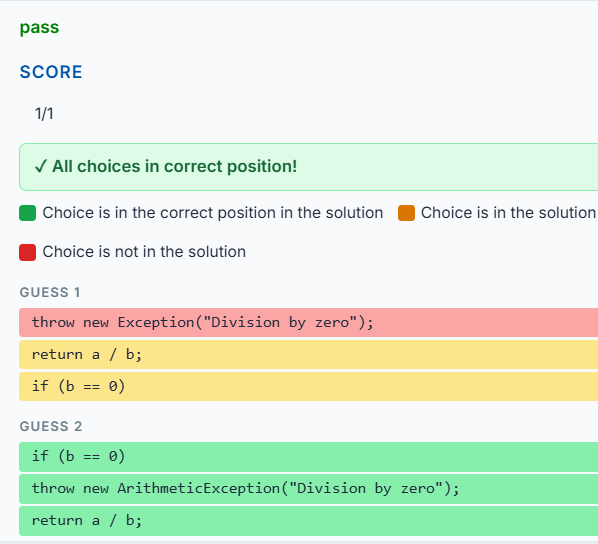}\vspace{3pt}\end{minipage}};%
  \draw[arr] (ps4.east) -- (ps5.west);%
  \draw[arr] (ps5.east) -- (ps6.west);%
  \node[panel] (al) at (1.85,-4.5) {\begin{minipage}[c]{2.9cm}\noindent\hspace{-4pt}{\setlength{\fboxsep}{0pt}\colorbox{black}{\parbox{\dimexpr\linewidth+8pt\relax}{\vspace{3pt}\hspace{4pt}\color{white}\scriptsize\bfseries G -- ACTIVITY LOG\vspace{3pt}}}}\\[3pt]\includegraphics[width=2.7cm,height=2.8cm,keepaspectratio]{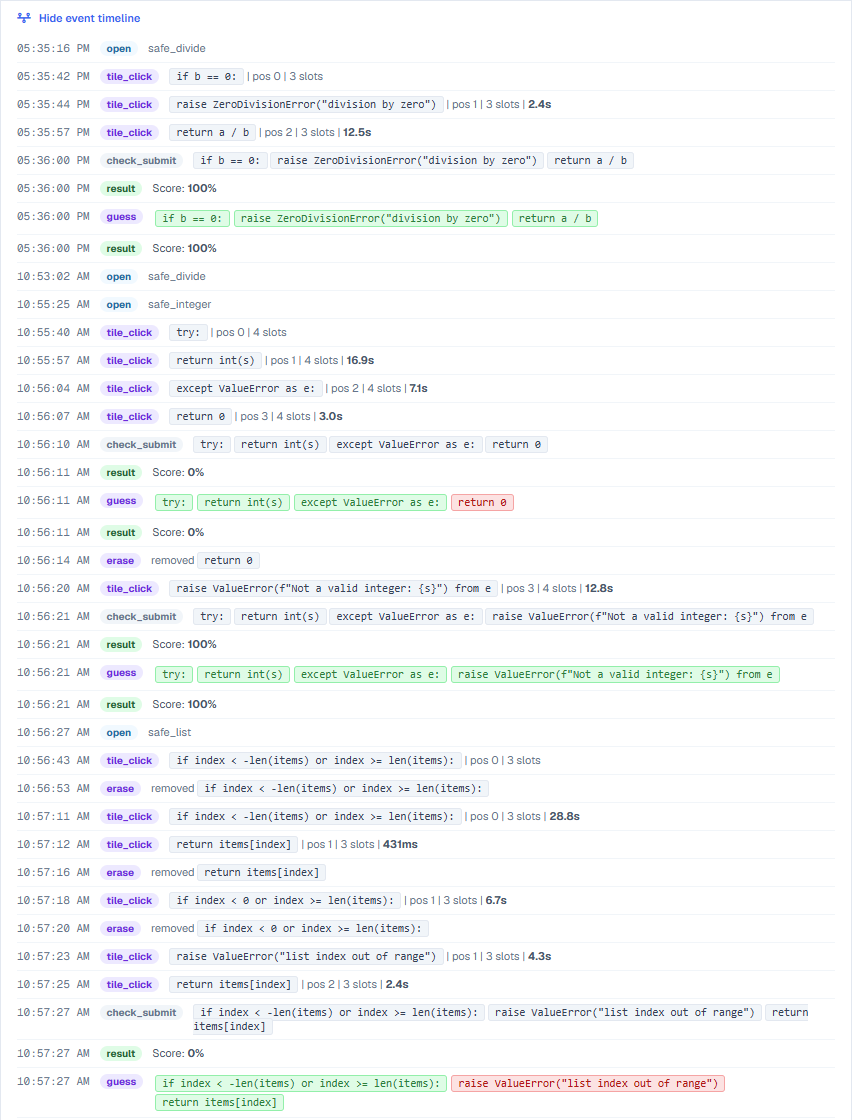}\vspace{3pt}\end{minipage}};%
  \node[panel,right=0.3cm of al] (lk) {\begin{minipage}[c]{2.9cm}\noindent\hspace{-4pt}{\setlength{\fboxsep}{0pt}\colorbox{black}{\parbox{\dimexpr\linewidth+8pt\relax}{\vspace{3pt}\hspace{4pt}\color{white}\scriptsize\bfseries H -- ATTEMPT HISTORY\vspace{3pt}}}}\\[3pt]\includegraphics[width=2.7cm,height=2.8cm,keepaspectratio]{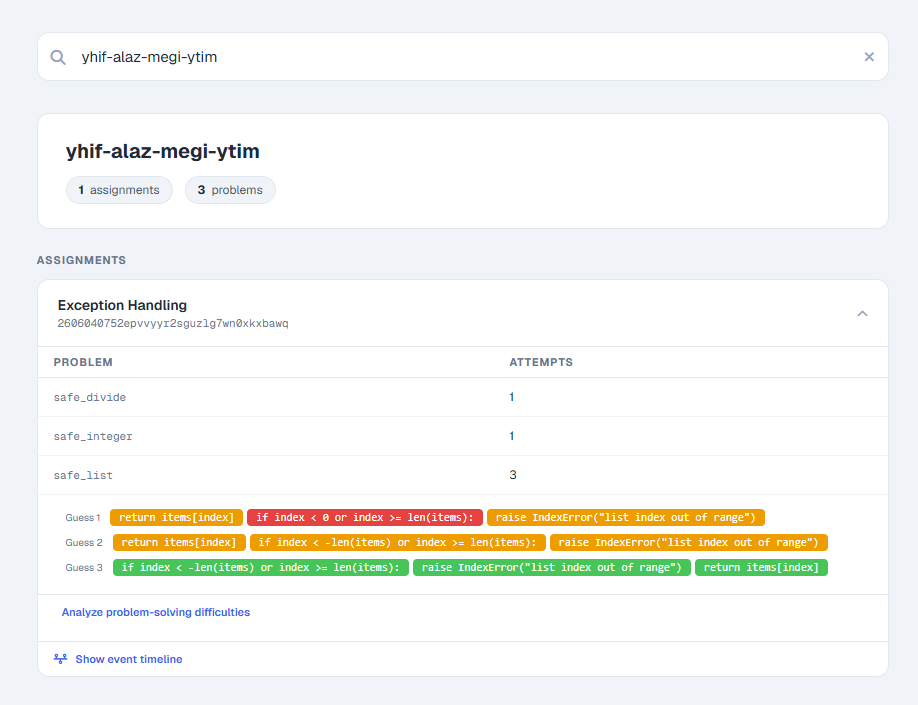}\vspace{3pt}\end{minipage}};%
  \node[panel,right=0.3cm of lk] (mc) {\begin{minipage}[c]{2.9cm}\noindent\hspace{-4pt}{\setlength{\fboxsep}{0pt}\colorbox{black}{\parbox{\dimexpr\linewidth+8pt\relax}{\vspace{3pt}\hspace{4pt}\color{white}\scriptsize\bfseries I -- DIFFICULTIES~{\small\faRobot}\vspace{3pt}}}}\\[5pt]%
       {\setlength{\fboxsep}{2pt}\colorbox{black}{\color{white}\tiny\bfseries Incorrect Exception Type}}\\[3pt]%
    {\setlength{\fboxsep}{2pt}\colorbox{black}{\color{white}\tiny\bfseries Wrong Control Flow Order}}\\[3pt]%
    {\setlength{\fboxsep}{2pt}\colorbox{black}{\color{white}\tiny\bfseries Using Return Instead of Throw}}\vspace{6pt}%

  \end{minipage}};%
  \node[panel,right=0.3cm of mc] (sl) {\begin{minipage}[c]{2.9cm}\noindent\hspace{-4pt}{\setlength{\fboxsep}{0pt}\colorbox{black}{\parbox{\dimexpr\linewidth+8pt\relax}{\vspace{3pt}\hspace{4pt}\color{white}\scriptsize\bfseries J -- CLASS SLIDES~{\small\faRobot}\vspace{3pt}}}}\\[2pt]\includegraphics[width=2.7cm,height=1.3cm,keepaspectratio]{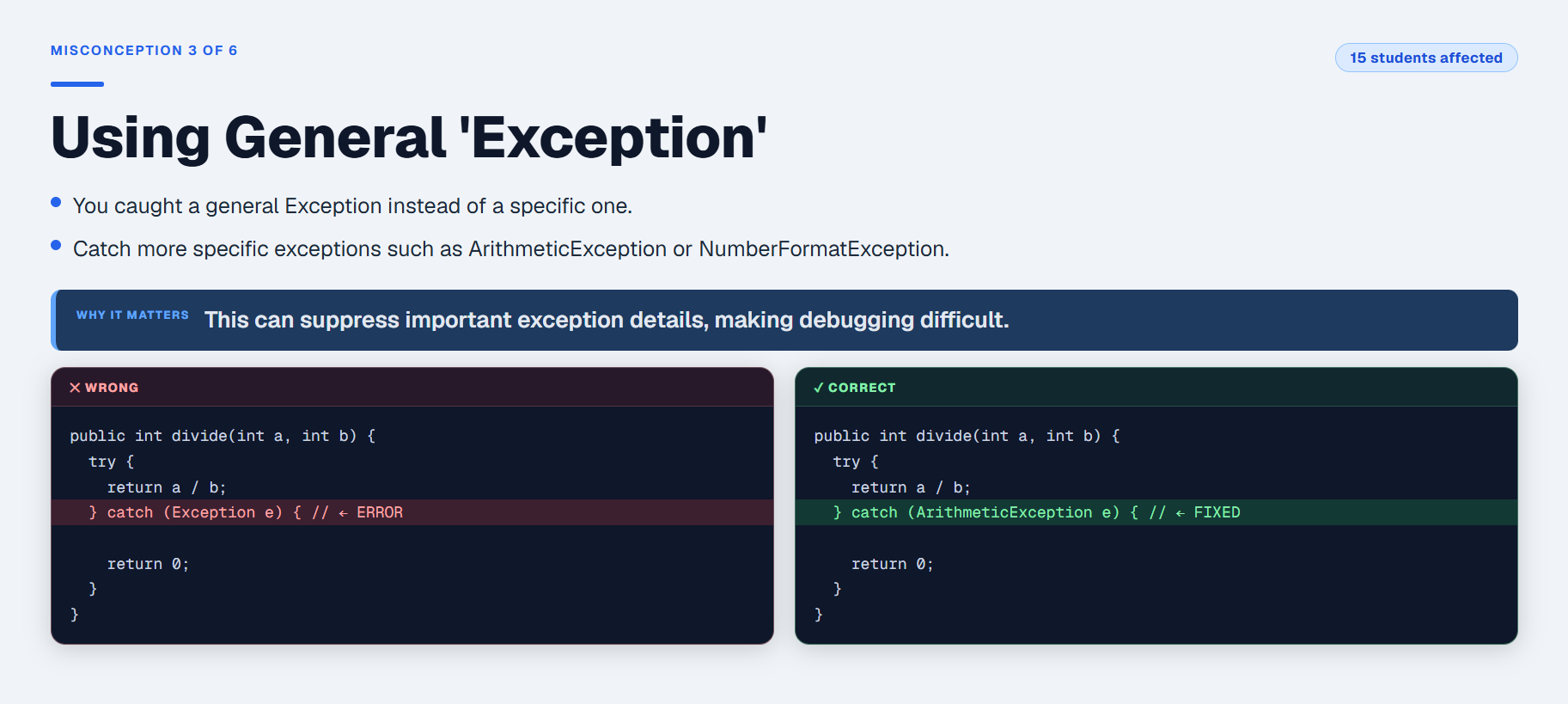}\\[2pt]\includegraphics[width=2.7cm,height=1.3cm,keepaspectratio]{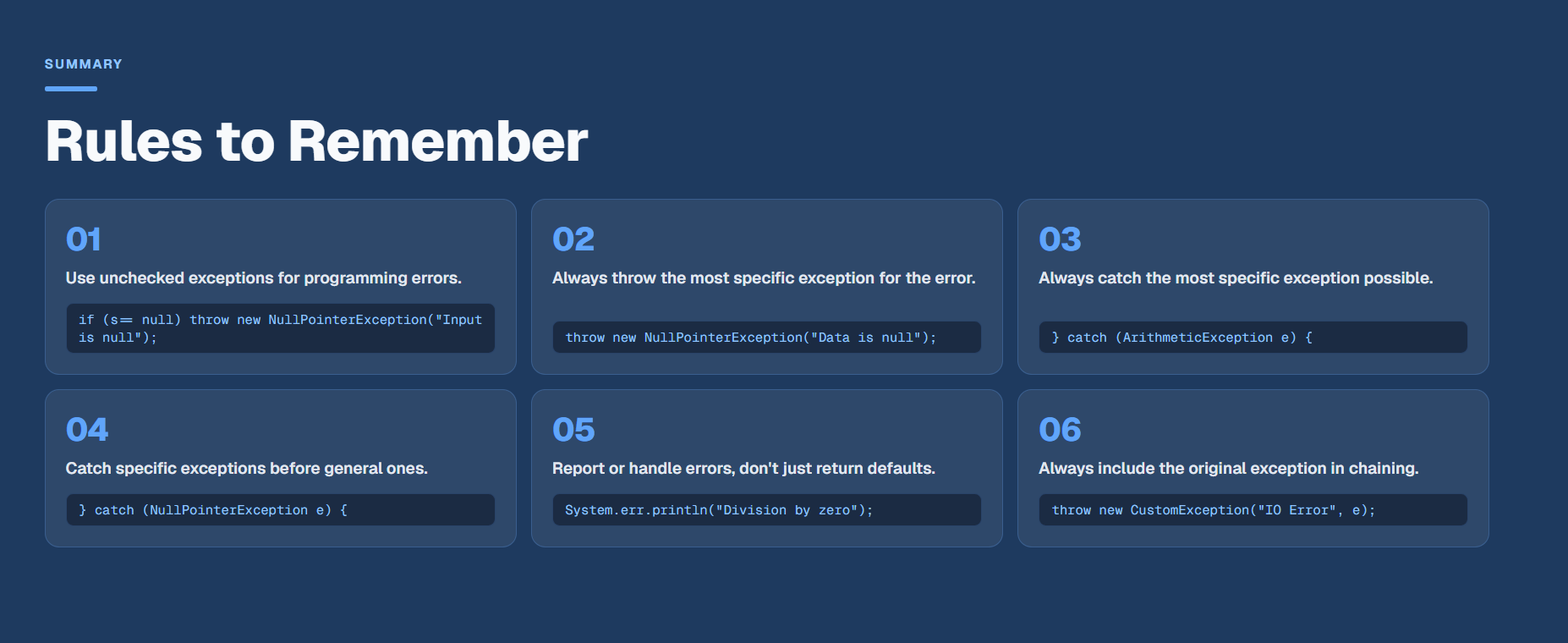}\vspace{3pt}\end{minipage}};%
  \node[draw=black!60, thin, fill=black, rounded corners=2pt, inner sep=5pt, font=\scriptsize\bfseries, text=white] (st) at (4.59,-2.0) {STUDENT PROFILE VIEW};%
  \node[draw=black!60, thin, fill=black, rounded corners=2pt, inner sep=5pt, font=\scriptsize\bfseries, text=white] (gs) at (sl |- st) {SUBMISSIONS};%
  \draw[-,thick,color=black] (ps6.south) -- ++(0,-0.5) coordinate (ps6branch);%
  \draw[arr] (ps6branch) -| (st.north);%
  \draw[arr] (ps6branch) -| (gs.north);%
  \draw[arr] (st.south) -- ++(0,-0.3) -| (al);%
  \draw[arr] (st.south) -- ++(0,-0.3) -| (lk);%
  \draw[arr] (lk.east) -- (mc.west);%
  \draw[arr] (gs.south) -- (sl.north);%
  \end{tikzpicture}
  \caption{Overview of workflow. Students solve Parsons problems (A--F); each block arrangement is recorded as a behavioral trace. Instructors use the Student Profile view to inspect per-student attempt histories~(H) and activity logs~(G). On demand, an AI pipeline analyzes
block arrangements to surface per-student problem-solving difficulty
patterns~(I) and generate class-level intervention slides~(J).}
  \Description{System overview with student and instructor panels.}
  \label{fig:overview}
\end{figure*}
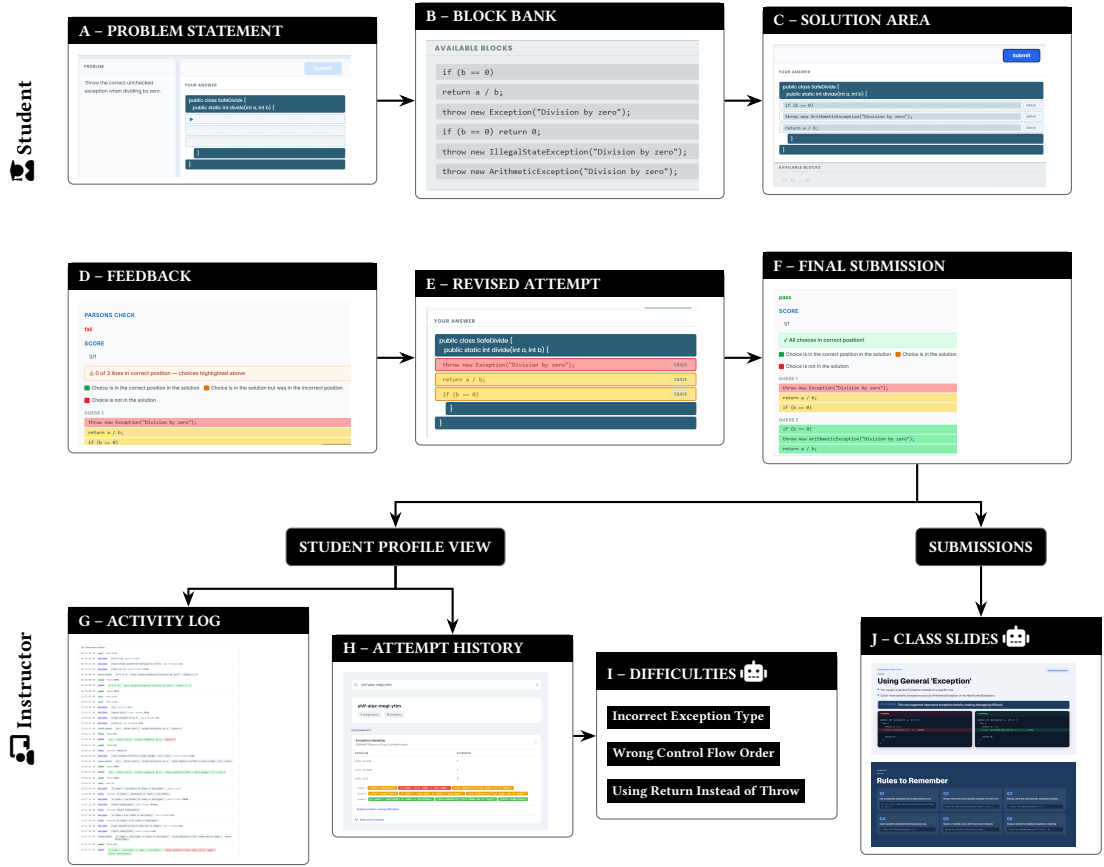

The tool extends CodeCheck~\cite{horstmann_codecheck}, an open-source web-based programming assessment platform, with a behavioral instrumentation layer designed to capture students' problem-solving processes while working on Parsons problems. 
Figure~\ref{fig:overview} illustrates the overall workflow of the tool.
Rather than recording only final submission outcomes, the tool captures fine-grained interaction events (Table~\ref{tab:events}) associated with each submission attempt, alongside the per-block 
feedback returned at each check. These events are persisted for subsequent analysis and together form a behavioral trace, a chronological record of the decisions students make before arriving at a final submission, states that remain invisible in on-demand feedback systems~\cite{helminen2012}. To support instructional decision-making, the tool provides a web-based instructor dashboard that transforms these traces into actionable insights.
Instructors can inspect per-student attempt histories and activity logs (panels G--H), optionally trigger AI-assisted analysis of problem-solving difficulties for individual students (panel I), and generate a summary of  common problem-solving difficulty patterns across the cohort as PowerPoint slides (panel J). 


\subsection{Design Rationale}

\subsubsection{Client-Side Event Capture}

The tool captures interaction events at the browser level. This design decision is what makes the intermediate states visible: each block placement, removal, and retry are recorded as it happens (Table~\ref{tab:events}), before the student decides to submit. Server-side instrumentation would capture only the final arrangement at each submission, losing the within-attempt behavior that is most diagnostic of conceptual difficulty.

\subsubsection{Human-in-the-Loop Problem-Solving Analysis}

\begin{figure}[t]
  \centering
  \begin{tikzpicture}
    \node[draw=black!25, fill=white, rounded corners=8pt, inner sep=6pt] (badge) {%
      \begin{tabular}{ccc}
        {\normalsize\faChalkboardTeacher} & {$\leftrightarrow$} & {\normalsize\faRobot}\\[2pt]
        {\small\textcolor{green!60!black}{\faThumbsUp}}~{\small\textcolor{red!70!black}{\faThumbsDown}} & & \\[2pt]
        \multicolumn{3}{c}{{\tiny\color{black!35}\faArrowDown}}\\[2pt]
        \multicolumn{3}{c}{{\normalsize\faUserGraduate}}
      \end{tabular}
    };
    \node[inner sep=0, right=0.4cm of badge] (img2) {%
  \includegraphics[width=0.68\linewidth]{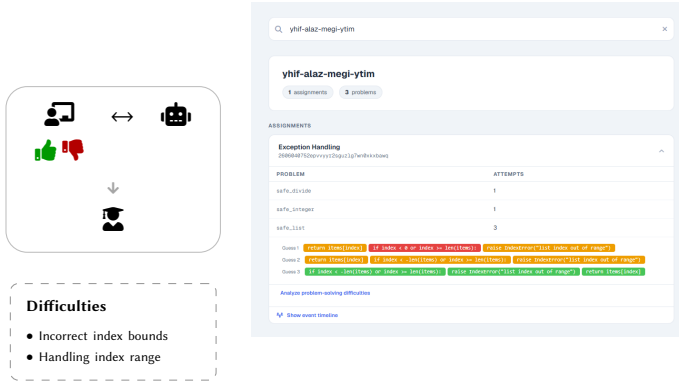}%
};
    \node[draw=black!40, dashed, fill=white, rounded corners=3pt,
      inner sep=6pt, font=\scriptsize\sffamily,
      text width=2.3cm, align=left,
      below=0.4cm of badge] (diff-lbl) {%
  \textbf{Difficulties}\\[4pt]
  {\tiny $\bullet$ Incorrect index bounds}\\[1pt]
  {\tiny $\bullet$ Handling index range}%
};
  \end{tikzpicture}
\caption{Human-in-the-loop review workflow.}
\label{fig:human-in-loop}

\end{figure}

Although the system could deliver AI-generated difficulty reports directly to students, the current deployment positions the instructor as a reviewer first. Figure~\ref{fig:human-in-loop} depicts this workflow: the instructor reviews the student's raw attempt history, optionally invokes the AI-assisted analysis to obtain a labeled difficulty report, and approves it before any feedback reaches the student, consistent with human-in-the-loop designs for LLM-generated educational feedback~\cite{soelch2026icse}. This reflects two concerns: AI-generated feedback can be technically correct yet unhelpful, since experts and students often disagree on hint quality~\cite{phung2025}; and we argue that instructors are better positioned than an automated system to judge whether a detected difficulty pattern is pedagogically meaningful given the course context and students' backgrounds.

\subsubsection{AI Analysis Pipeline}

When an instructor triggers the analysis for a student, the system serializes the student's complete attempt history for the selected assignment: each submission is represented as an ordered sequence of block identifiers paired with their per-block color-coded feedback (green, yellow, or red, as described in Section~\ref{sec:parsons-interface}). This structured trace is passed to GPT-4.1-mini via the OpenAI API, with a prompt that instructs the model to identify recurring conceptual difficulties evident in the pattern of incorrect placements and distractor selections across attempts. The model returns a short list of labeled difficulty patterns, each with a brief natural-language explanation, displayed to the instructor as the difficulty report (panel~I, Figure~\ref{fig:overview}). For class-level slides~(panel~J), a separate two-stage pipeline first extracts a small set of recurring difficulties from a sample of student traces, then classifies each student's attempt history against that shared list before aggregating the results into a class-wide summary. The analysis is on-demand: the pipeline runs only when an instructor explicitly invokes it from the Student Profile View or submissions page, and its output is reviewed by the instructor before any feedback reaches the student. The same model was used throughout both deployments.

\subsection{Parsons Problem Interface}
\label{sec:parsons-interface}

Students access Parsons problems through a link in the course LMS, which opens a dedicated web-based interface. Upon opening an assignment, students are presented with a problem statement and a bank of shuffled code blocks, including distractors (Figure~\ref{fig:overview}, panels A--B). Students arrange blocks into answer blocks by clicking blocks from the bank into the solution area (panel C). Unlike systems where indentation must be set manually~\cite{helminen2012}, our tool aligns blocks automatically upon placement, avoiding by design a source of difficulty documented in prior work: Helminen et al.\ found that in one assignment, three distinct misindentation patterns together accounted for 83\% of solutions. Once all blocks are filled, students submit their arrangement to receive immediate color-coded visual feedback analogous to the Wordle game: each slot is marked green if the block is in the correct position, yellow if the block belongs somewhere else in the solution, and red if the block is a distractor that does not appear in the correct solution at all (panel~D). Unlike a binary correct/incorrect judgment, this per-block feedback encodes both presence and position simultaneously, allowing students to identify which blocks need repositioning. Crucially, each submission together with its color-coded result is recorded as part of the student's behavioral trace (panel~H), making visible the reasoning path that produced the final submission. Students can then revise their arrangement and submit again (panels E--F).

\subsection{Instructor Interface}

From any assignment, instructors open the Student Profile View (Figure~\ref{fig:overview}, panels G--H) to search for a specific student and inspect their full attempt history and activity log. Instructors can use this view to prepare targeted feedback for students who may need additional support before office hours or tutoring sessions. From the same view, instructors can trigger an automated analysis of problem-solving difficulties that processes the student's attempt history and returns a set of identified problem-solving difficulty patterns (panel~I, Figure~\ref{fig:overview}), providing a structured starting point for the conversation. From the general submissions view, instructors can generate a class-level slide deck summarizing common problem-solving difficulty patterns across the cohort (panel~J, Figure~\ref{fig:overview}).

\subsection{Recorded Data}

Both the activity log and attempt history are persisted to a PostgreSQL database. Each row records an event type, a timestamp, and a JSON payload. Interaction events (block placements, removals, submission checks, and problem navigation) form the activity log; Table~\ref{tab:events} lists the six event types captured. Submission snapshots, each recording the complete block arrangement and per-block color-coded feedback, form the attempt history. The tool records identifiers as provided by the course platform: at Aalto University, the platform assigns randomly generated pseudonyms natively; at the University of Houston, student identifiers were pseudonymized prior to analysis.

\begin{table}[h]
\caption{Interaction events recorded in the activity log}
\label{tab:events}
\small
\begin{tabular}{@{}l p{5.2cm}@{}}
\toprule
\textbf{Event type} & \textbf{Description} \\
\midrule
problem\_open        & Student navigates to a problem \\
parsons\_tile\_click & Student places a block from the bank into the solution area \\
parsons\_erase       & Student removes a block from the solution area back to the bank \\
check\_submit        & Student submits the current block arrangement for evaluation \\
check\_result        & Score returned for the submitted arrangement \\
parsons\_guess       & Complete block arrangement at submission time with per-block color-coded feedback (green\,/\,yellow\,/\,red); stored as a submission snapshot forming the attempt history \\
\bottomrule
\end{tabular}
\smallskip

{\footnotesize The activity log displays abbreviated labels~(panel~G, Figure~\ref{fig:overview}); the names above reflect the full event types as stored in the database.}
\end{table}

\section{Classroom Deployment}

\subsection{Participants}

The system has been deployed in two universities and has been used in two courses with different course levels, programming languages, instructional formats, and student demographics. Table~\ref{tab:deployments} summarizes the key characteristics of each deployment. Together, these deployments allow us to examine whether the behavioral patterns we observe generalize across different student populations and instructional contexts, or whether they are specific to a particular course context. Although two deployments are insufficient to support strong generalizability claims, the diversity of contexts increases confidence that the observed patterns and lessons learned are not unique to a single course setting.

\begin{table}[t]
\caption{Deployment contexts for the two university courses.}
\label{tab:deployments}
\small
\begin{tabular}{@{}lll@{}}
\toprule
                        & \textbf{University of Houston}     & \textbf{Aalto University} \\
\midrule
Country                 & United States             & Finland               \\
Course level            & Upper-division Software Design           & CS1                   \\
Student population      & CS majors                 & non-CS majors         \\
Language                & Java                      & Python                \\
Format                  & In-person                 & Online, self-paced    \\
Semester                & Spring 2026               & Summer 2026           \\
Students                & 68                        & 36                   \\
Authentication          & Institutional SSO         & Anonymized            \\
\bottomrule
\end{tabular}
\end{table}

\subsubsection{University of Houston}

The first deployment was an upper-division Java course on software design at the University of Houston, a large public Hispanic and minority-serving research university in the United States. The study was conducted during Spring 2026 and involved 68 students. Students accessed assignments through the institution's LMS using university credentials, enabling interaction traces to be linked to student identifiers across assignments.

\subsubsection{Aalto University}

The second deployment was an introductory Python course for non-CS majors at Aalto University in Finland, conducted online with students progressing at their own pace. The study was conducted during Summer 2026 and involved 36 students. Students accessed assignments via a direct link without institutional authentication; traces were recorded under anonymized identifiers for data privacy.

\begin{table}[t]
\caption{Parsons problems deployed in both courses.}
\label{tab:problems}
\resizebox{\columnwidth}{!}{%
\begin{tabular}{@{}l l l p{3.6cm}@{}}
\toprule
\textbf{ID} & \textbf{Univ.} & \textbf{Bl./Dist.} & \textbf{Description} \\
\midrule
P1 Safe Divide  & UH \& Aalto & 3\,/\,3 & Raise/throw correct exception for division by zero \\
P2 Safe Integer & UH \& Aalto & 4\,/\,3 & Raise/throw correct exception for invalid string input \\
P3$^\dagger$ Checked/Unchecked & UH & 3\,/\,2 & Choose correct exception type per scenario \\
P3$^\dagger$ Safe List Access & Aalto & 3\,/\,3 & Raise correct exception for out-of-range index \\
\bottomrule
\end{tabular}}
\smallskip

{\footnotesize UH\,=\,University of Houston; Aalto\,=\,Aalto University.
Bl./Dist.\ = blocks in correct solution\,/\,distractor blocks included.
$^\dagger$P3 problems share the exception-handling theme but differ in scenario across deployments.}
\end{table}

\begin{table*}[t]
\centering
\caption{Problem difficulty and error patterns by deployment.
  \textit{n} = students who submitted;
  Avg.\ Att.\ = mean block-arrangement attempts;
  1st~Att.\ = correct on first attempt;
  Incorr.\ = \% of attempts containing at least one wrong block.
  Error-pattern percentages are computed over incorrect attempts only
  (Err.\ \textit{n} = students with $\geq$1 such attempt).}
\label{tab:combined}
\small
\begin{tabular}{ll r rrrr r rrr}
\toprule
& & & \multicolumn{4}{c}{\textit{Submissions}}
  & \phantom{x}
  & \multicolumn{3}{c}{\textit{Error patterns}} \\
\cmidrule(lr){4-7}\cmidrule(l){9-11}
\textbf{Univ.} & \textbf{Assign.} & \textit{n}
  & \textbf{Avg.\ Att.} & \textbf{1st Att.}
  & \textbf{Err.\ \textit{n}} & \textbf{Incorr.}
  &
  & \textbf{Exc.\ type} & \textbf{Ret.\ not raise}
  & \textbf{Wrong CF} \\
\midrule
UH & P1          & 68 & 1.91 & 67.6\%  & 30 & 44\% && 90\% & 60\% & 43\% \\
UH & P2          & 67 & 1.93 & 68.7\%  & 25 & 37\% && 48\% & 52\% & 52\% \\
UH & P3$^\dagger$& 66 & 2.62 & 33.3\%  & 48 & 73\% && 37\% & 47\% & 43\% \\
\midrule
Aalto & P1 & 36 & 1.25 & 80.6\% &  7 & 19\% &&  57\% & 43\% &  0\% \\
Aalto & P2 & 35 & 1.69 & 54.3\% & 16 & 46\% &&  62\% & 50\% & 31\% \\
Aalto & P3$^\dagger$& 35 & 1.91 & 22.9\% & 27 & 77\% &&  89\% & 30\% & 15\% \\
\bottomrule
\end{tabular}
\smallskip\\
\raggedright\footnotesize
UH\,=\,University of Houston; Aalto\,=\,Aalto University.
P1\,=\,Safe Divide; P2\,=\,Safe Integer;
P3\,=\,Checked vs.\ Unchecked~(UH)\,/\,Safe List Access~(Aalto).
$^\dagger$Hardest problem in each deployment; both target
exception-raising vs.\ silent exit with language-appropriate
syntax (Java\,/\,Python).
Exc.\ type = exception type confusion;
Ret.\ not raise = \texttt{return} value instead of raising
an exception;
Wrong CF = wrong control-flow structure (guard condition or
block ordering).
\end{table*}

\subsection{Task Design}

Both courses included exception handling as a common topic, making it a natural choice for cross-deployment comparison. Other assignment types deployed at the University of Houston, such as design patterns, were not part of the Aalto University curriculum and therefore could not be compared across cohorts. The assignment consisted of three Parsons problems targeting exception handling concepts common to both Java and Python, adapted to the syntax of each language. Although the scenarios differed across courses, both versions targeted the same learning objective: selecting the semantically appropriate exception type for a given error condition. In both deployments, problems were presented sequentially (P1--P3), though students could advance without solving the current problem. P1 and P2 were deployed identically across both universities, adapted only to the syntax of each language; P3 differed in scenario across deployments. Table~\ref{tab:problems} summarizes the three problems.

\section{Results}

\subsection{Difficulties Encountered by Students}
\subsubsection{Return Instead of Raise or Throw}
\label{sec:return-throw}

A recurring difficulty across both deployments was substituting a \texttt{return} statement for an exception throw. At the University of Houston ($n = 68$), 60\% of students with incorrect attempts placed \texttt{if (b == 0) return 0} in SafeDivide~(P1) rather than \texttt{throw new ArithmeticException(\ldots)}, and 47\% substituted \texttt{return} for \texttt{throw} in CheckedVsUnchecked~(P3). At Aalto University, this pattern was prominent in \texttt{safe\_integer}~(P2), where 50\% of students with incorrect attempts substituted \texttt{return} for \texttt{raise ValueError}, treating a parse failure as a silent exit rather than an exceptional event. In both cases, students treated the error condition as a reason to exit early rather than to signal it as an exception. This pattern coincided with the lowest first-attempt accuracy at the University of Houston: 33.3\% on CheckedVsUnchecked, compared to 67--69\% on the other two problems (Table~\ref{tab:combined}).

\subsubsection{Wrong Exception Type Selection}
\label{sec:exception-type}

Both deployments revealed a consistent pattern of incorrect exception type selection. At the University of Houston, exception type confusion was the dominant pattern in P1, affecting 90\% of students with incorrect attempts, and remained present in P2 (48\%) and P3 (37\%): in each case students placed a semantically imprecise type, such as \texttt{IllegalStateException}, \texttt{Exception}, or \texttt{RuntimeException}, where a more specific subtype was required. At Aalto University, exception type confusion was present across both harder problems: 62\% of students with incorrect attempts in \texttt{safe\_integer}~(P2) involved a broad exception class (\texttt{except Exception}) or wrong exception type (\texttt{raise RuntimeError}); in \texttt{safe\_list}~(P3), most students with incorrect attempts (89\%) selected an incorrect guard condition. The dominant error in P3 was choosing \texttt{if index < 0} rather than the correct bound \texttt{if index < -len(items)}, which is required to reject out-of-range negative indices. This reflects confusion about \emph{when} to raise rather than \emph{which} exception to raise, which may indicate a subtler conceptual gap than selecting the wrong exception class.

\subsubsection{Incorrect Control-Flow Ordering}
\label{sec:control-flow}

Control-flow ordering errors appeared consistently across all three problems at the University of Houston (43--52\% of students with incorrect attempts), but without a dominant pattern: in P3, wrong ordering (43\%) was roughly co-equal with return-for-raise substitution (47\%) and exception type confusion (37\%), suggesting upper-division students cycled through all three difficulty types rather than concentrating on one. At Aalto University, control-flow ordering errors were a secondary pattern in \texttt{safe\_integer}~(P2), present in 31\% of students with incorrect attempts; the dominant difficulties were exception type confusion (62\%) and return-for-raise substitution (50\%). In \texttt{safe\_list}~(P3), ordering errors were present in 15\% of students with incorrect attempts; instead, most students selected a semantically incorrect guard condition (a distractor checking \texttt{index~<~0} rather than \texttt{index~<~-len(items)}), captured by the exception-type column in Table~\ref{tab:combined} (89\%). One plausible explanation is that the conceptual barrier shifts depending on problem structure: when the block arrangement pattern is unfamiliar (try/except sequencing), students cycle through orderings; when the structure is intuitive, the difficulty moves to semantic precision. However, this pattern is confounded with the language, course level, and population differences between deployments.

\subsection{Individual Trajectories}
\label{sec:loops}

\begin{figure}[t]
  \centering
  \includegraphics[width=0.85\linewidth]{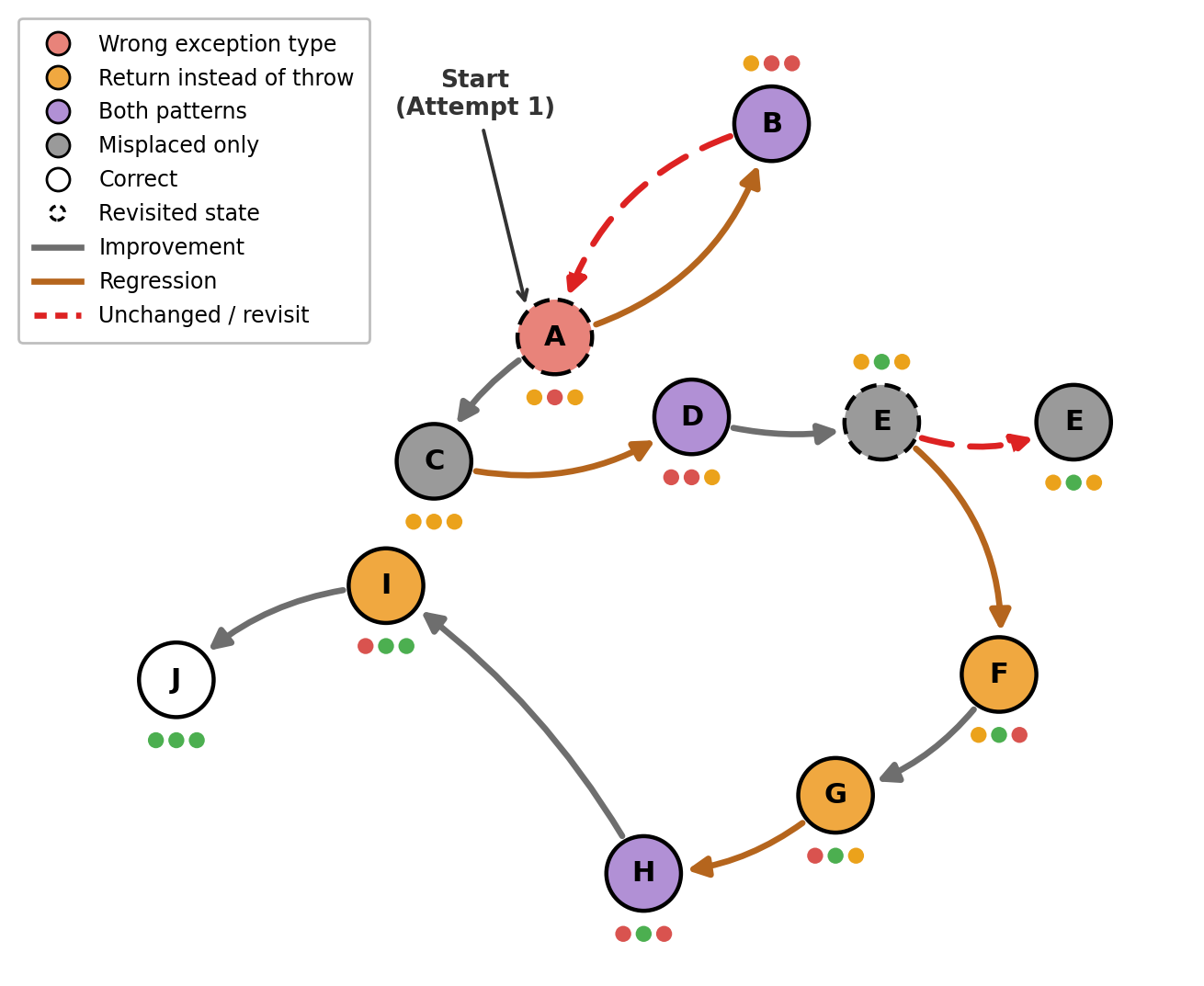}
  \caption{State-transition map of one University of Houston student's 12-attempt sequence on P3.}
  \label{fig:state-transition}
\end{figure}

Visualizing a solving session as a graph exposes patterns and anomalies. Aggregate counts (Table~\ref{tab:combined}) show how often each difficulty pattern occurred, but not how a student moved between them within a single attempt sequence. Figure~\ref{fig:state-transition} traces one University of Houston student's twelve submissions on P3 as a state-transition graph, where each node is a distinct 3-block arrangement and arrows show the order attempts were submitted. Node fill follows the same categories as Table~\ref{tab:combined}: red marks the wrong exception type, orange marks \texttt{return} substituted for \texttt{throw}, purple marks both at once, grey marks no wrong blocks but at least one misplaced, and white marks the correct arrangement; a dashed border (A,~E) marks a state the student returned to more than once. Arrow color marks whether the next attempt improved on, regressed from (brown), or exactly repeated (red dashed) the one before it. The student's path is non-monotonic: after an initial detour back to an earlier arrangement (A$\to$B$\to$A) and an unchanged resubmission (E$\to$E), the trace alternates between the two difficulty patterns before converging on the correct arrangement. The full trace reveals four regressions and two unproductive resubmissions across twelve attempts. Across the University of Houston, between 6\% and 12\% of solving sessions contained at least one state revisit, below the 21--33\% reported by Helminen et al.~\cite{helminen2012} for similar Parsons tasks; at Aalto University, revisits were rare (0--3\% across problems) and absent in P3, the problem from which Figure~\ref{fig:state-transition} is drawn.
 
\subsection{Instructor Perspectives}

Instructors from both deployments shared their experience using the dashboard during the course:

\begin{tcolorbox}[
  breakable,
  colback=gray!8, colframe=gray!30,
  arc=2pt, boxrule=0.4pt,
  left=5pt, right=5pt, top=2pt, bottom=2pt
]
\footnotesize\itshape
``I liked the intuitive interface and the analysis that it provided. Designing new problems was hassle-free. I could use bite-size Parson problems as an in-class exercise that students could do in short amount of time. The summarizing slides were helpful to explain the common mistakes that students make right there in the class. It was also great that they could not copy the solutions from AI, paste the solution in the editor and submit. I felt that I have a quick good estimate of their skill levels.'' \hfill\upshape\normalfont\footnotesize ---Instructor, University of Houston
\end{tcolorbox}
\begin{tcolorbox}[
  breakable,
  colback=gray!8, colframe=gray!30,
  arc=2pt, boxrule=0.4pt,
  left=5pt, right=5pt, top=2pt, bottom=2pt
]
\footnotesize\itshape
``From an instructor perspective, the dashboard was useful because it made students' intermediate solution attempts visible rather than showing only the final outcome. The color-coded attempt history made it easy to see which lines were correct, which were merely out of place, and which were incorrect distractors, helping me identify whether a student was close to the solution or repeatedly applying the same misconception. I found the AI-assisted analysis useful as a starting point for interpreting these patterns, but I would still want to review the traces of students before deciding what feedback or intervention is appropriate.'' \hfill\upshape\normalfont\footnotesize ---Instructor, Aalto University
\end{tcolorbox}

\section{Discussion, Limitations, and Future Work}

The overall aim of this work was to make the problem-solving processes of the students more visible to the instructors, rather than merely their final submissions. The deployment demonstrates that this is achievable in practice: the same event-capture and AI-analysis pipeline operated without modification across two programming languages, two course levels, and two institutions, revealing similar recurring difficulty patterns among CS1 students and upper-division majors (Table~\ref{tab:combined}). However, the extent to which this visualization can be most effective is an area that requires more investigation~\cite{pahi2026inclass, zamfirescupereira2025bot, soelch2026icse}.

We have anecdotal evidence that class-level aggregation (panel~J, Figure~\ref{fig:overview}) is particularly valuable in practice, since identifying recurring difficulties across an entire cohort is impractical through manual inspection of individual students' work alone. 
At the same time, both instructors supplemented the AI-generated reports by reviewing students' interaction traces before deciding how to provide feedback. 

Beyond its use in the classroom, the tool also has potential as a research instrument by enabling large-scale analyses of programming behavior that would otherwise be very time-consuming. Realizing that potential, however, will depend on wider adoption; we intend to release the complete implementation to facilitate future use and replication.


\paragraph{Limitations.} Three limitations apply to these findings. The comparison was constrained to exception handling, the only topic common to both curricula, and may not generalize to other domains; the Aalto University cohort was small ($n = 36$), limiting cross-deployment weight; the dashboard's value depends on instructor engagement.

\paragraph{Future Work.} Five extensions follow directly from the findings. First, isomorphic variants~\cite{tilantera2025traces} would separate understanding from accidental correctness. Second, automated puzzles from recurring distractor selections~\cite{hou2024codetailor} would target these patterns directly. Third, real-time alerts on repeated failed arrangements~\cite{price2017isnap} would enable earlier intervention. Fourth, an in-dashboard state-transition graph would surface cycles and regressions automatically. Lastly, analyzing the quality of the AI-generated reports about common errors systematically would be important for quality assurance.

\section{Acknowledgments}
This material is based upon work supported by the U.S. National Science Foundation under Grant No. 2225373. Any opinions, findings,
and conclusions or recommendations expressed in this material are
those of the author(s) and do not necessarily reflect the views of
the National Science Foundation.


\bibliographystyle{ACM-Reference-Format}
\bibliography{references}


\end{document}